\documentclass[11pt]{article}
\usepackage{epsfig}

\def\la{\langle}\def\ra{\rangle}
\def\be{\begin{eqnarray}}\def\bea{\begin{eqnarray}}
\def\ba{\begin{eqnarray}}
\def\ee{\end{eqnarray}}\def\eea{\end{eqnarray}}
\def\ea{\end{eqnarray}}
\def\ben{\begin{enumerate}}\def\bitem{\begin{itemize}}
\def\een{\end{enumerate}}\def\eitem{\end{itemize}}\def\bi{\bibitem}
\def\prl{Phys. Rev. Lett.}\def\np{Nucl. Phys.}

\begin{document}
\begin{titlepage}
\hfill {KIAS-P03026}

\hfill {\today: hep-ph/}

\begin{center}
\ \\
{\Large \bf  Sliding Vacua in Dense Skyrmion Matter}
\\
\vspace{.30cm}

Hee-Jung Lee$^{a,b}$, Byung-Yoon Park$^{a,b}$, Mannque Rho$^{a,c}$
\\ and Vicente Vento$^{a,d}$

\vskip 0.20cm

{(a) \it School of Physics, Korea Institute for Advanced Study,
Seoul 130-722, Korea}

{(b) \it Department of Physics,
Chungnam National University, Daejon 305-764, Korea}\\
({\small E-mail: hjlee@phya.snu.ac.kr,
bypark@chaosphys.cnu.ac.kr})


{(c) \it Service de Physique Th\'eorique, CE Saclay}\\
{\it 91191 Gif-sur-Yvette, France}\\
({\small E-mail: rho@spht.saclay.cea.fr})

{(d) \it Departament de Fisica Te\`orica and Institut de
F\'{\i}sica
Corpuscular}\\
{\it Universitat de Val\`encia and Consejo Superior
de Investigaciones Cient\'{\i}ficas}\\
{\it E-46100 Burjassot (Val\`encia), Spain} \\ ({\small E-mail:
Vicente.Vento@uv.es})

\end{center}
\vskip 0.3cm

\centerline{\bf Abstract}

\vskip 0.5cm

In continuation of our systematic effort to understand hadronic
matter at high density, we study dense skyrmion matter and its
chiral phase structure in an effective field theory implemented
with the trace anomaly of QCD applicable in the large $N_c$ limit.
By incorporating a dilaton field $\chi$ associated with broken
conformal symmetry of QCD into the simplest form of skyrmion
Lagrangian, we simulate the effect of ``sliding vacua" influenced
by the presence of matter and obtain what could correspond to the
``intrinsic dependence" on the background of the system, i.e.,
matter density or temperature, that results when a generic chiral
effective field theory of strong interactions is matched to QCD at
a matching scale near the chiral scale $\Lambda_\chi \sim 4\pi
f_\pi\sim 1$ GeV. The properties of the Goldstone pions and the
dilaton scalar near the chiral phase transition are studied by
looking at the pertinent excitations of given quantum numbers on
top of a skyrmion matter and their behavior in the vicinity of the
phase transition from Goldstone mode to Wigner mode characterized
by the changeover from the FCC crystal to the half-skyrmion CC
crystal. We recover from the model certain features that are
connected to Brown-Rho scaling and that suggest how to give a
precise meaning to the latter in the framework of an effective
field theory that is matched to QCD .

\vskip 0.3cm \leftline{Pacs: 12.39-x, 13.60.Hb, 14.65-q, 14.70Dj}
\leftline{Keywords: chiral symmetry, dilaton, pion, dense matter,
mended symmetry, BR scaling}

\end{titlepage}

\section{Introduction}

In trying to understand what happens to hadrons under extreme
conditions, e.g., at high density as in compact stars or at high
temperature as in relativistic heavy-ion collisions, it is
necessary that the theory or model adopted for the description be
consistent with the basic tenet of QCD. What this means in terms
of effective theories using macroscopic degrees of freedom, i.e.,
hadrons, the effective field theory should be matched to QCD at a
scale close to the chiral scale $\Lambda_\chi \sim 4\pi f_\pi\sim
1$ GeV. In a recent important development reviewed in
\cite{HY:PR}, Harada and Yamawaki show how this matching can be
effectuated in the framework of hidden local symmetry (HLS)
theory. In this theory, the light-quark vector meson fields
$\rho_\mu$ figure as gauge fields coupled to the Goldstone pion
fields $\pi$. When this theory is matched to QCD via current
correlators at a suitable scale $\Lambda_M\sim \Lambda_\chi$,
among the variety of flow paths by which the theory can follow to
its multiple fixed points as the scale is changed, it picks the
particular flow that leads to one unique fixed point as one
reaches the point at which the phase change to Wigner mode from
Goldstone mode takes place. This fixed point corresponds to the
``vector manifestation (VM)" in which the $parameters$ of the
Lagrangian have the limiting behavior:
 \be
g\rightarrow 0, \ \ \ a\rightarrow 1\label{VM}
 \ee
where the parametric $g$ is the (hidden) gauge coupling constant
and $a$ is the ratio $F_\pi/F_\sigma$ where $F_\pi$ and $F_\sigma$
are respectively the $parametric$ pion~\footnote{We will denote
the physical pion decay constant by the lower case $f_\pi$ to be
distinguished from the parametric one.} and would-be Goldstone
scalar decay decay constants. In \cite{HY:VM}, Harada and Yamawaki
discuss how the fixed point (\ref{VM}) is reached when one dials
the number of flavors to a critical value $N_f^c > 3$ at which the
physical pion decay constant $f_\pi$ vanishes. Subsequently it has
been shown that chiral phase transition at critical temperature
$T_c$~\cite{HS} and at critical density $n_c$~\cite{HKR} involves
the same VM. The consequence is that at the critical point,
independently of whether it is driven by temperature, density or
large $N_f$, the gauge boson mass (both parametric and pole)
approaches zero and so does the constituent quark mass. Thus all
light-quark hadrons (except for the Goldstone pions) are expected
to become massless in the chiral limit in some power of the gauge
coupling constant $g$. This observation led Brown and Rho to
conjecture that the property of VM and Brown-Rho scaling are
intimately connected~\cite{BR:Berkeley}. An important lesson one
can learn from this development is that without the Wilsonian
matching to QCD, there is no way to pick the right one from the
multitude of different ways to describe hadron properties near
chiral restoration. This means that the theories that possess all
relevant symmetries but are unmatched to QCD have no predictive
power for hadron properties. This caveat applies not only to HLS
theory but also to all chiral field theories with or without
vector-meson degrees of freedom.

The matching at $\Lambda_M$ of the correlators between the HLS
sector and the QCD sector gives the parameters of the ``bare" HLS
Lagrangian in terms of the quantities figuring in the QCD sector,
namely, the color gauge coupling $\alpha_s$, $\Lambda_{QCD}$ and
quark and gluon condensates. In medium, the condensates depend on
the background, that is, on density~\footnote{Similar arguments
hold in heat bath although we do not specifically mention it in
what follows.}. Imposing that the quark condensate
$\la\bar{q}q\ra$ -- as an order parameter -- vanish at the
critical density $n_c$, one obtains, at the matching scale, the
conditions on the HLS parameters $\ g(\Lambda_M,n_c)=0$ and
$a(\Lambda_M,n_c)=1$. However it gives no constraint on the
parametric $F_\pi (\Lambda_M,n_c)$. Now bringing the scale from
$\Lambda_M$ to the scale appropriate to the physics involved is
done by renormalization group equations. Thus the properties both
at the matching scale in matter-free vacuum and at the VM in
medium are completely determined by the theory. But the HLS theory
formalism does not offer any method to describe how the parameters
behave at an arbitrary scale $\Lambda$ and density $n$ away from
the two special points. This means that we know practically
nothing on the $intrinsic$ density dependence of the parameters of
the Lagrangian and hence cannot compute the density dependence of
physical quantities except at $n\sim 0$ and $n\sim n_c$. Thus it
would be futile to attempt to employ HLS/VM theory in its present
form for analyzing experimental data which of course sample all
ranges of density.

In this paper, we tackle the {\it intrinsic dependence} and
related issues with the help of the unified approach to high
density developed in \cite{LPMRV03}. In \cite{LPMRV03}, it was
shown using a skyrmion Lagrangian in its simplest form that an
effective Lagrangian valid in the large $N_c$ limit (where $N_c$
is the number of colors) can describe both infinite nucleonic
matter and pionic fluctuations thereon as density is increased and
that by describing the chiral restoration as a changeover from an
FCC crystal to the half-skyrmion matter, one can describe the
properties of various excitations commensurate with the background
given by the half-skyrmion crystal configuration. In this
approach, the density dependence of the background is taken into
account to all orders. No low-density approximation whose validity
is in doubt except at very low density is ever made in the
calculation. The power of the approach is that the dynamics of the
background and excitations thereon can be treated in a unified way
on the same footing with a single Lagrangian.

In addressing the problem at hand, the Lagrangian used in
\cite{LPMRV03} is probably incomplete. In fact, it is not clear
that the intrinsic density dependence required by the matching to
QCD discussed above is fully implemented in the model. One
puzzling feature we found in \cite{LPMRV03} was that the Wigner
phase represented by the half-skyrmion matter with $\la {\rm tr}
U_0\ra=0$ supported a non-vanishing pion decay constant. This was
interpreted there as a possible signal for a pseudo-gap phase.
However it can also be interpreted as an analog to Georgi's
``vector limit" in which chiral symmetry is restored with pions
present with non-vanishing pion decay constant. Harada and
Yamawaki argue~\cite{HY:PR} that this phase is not consistent with
QCD Ward identity, implying that it cannot be realized in nature.

Our proposal in this paper is that we can circumvent the above
difficulty and achieve our objective within a large $N_c$
framework by incorporating into the skyrmion description the trace
anomaly of QCD which is lacking in the standard skyrmion model. In
fact it is this feature that led to BR scaling in \cite{BR91} when
the problem was treated albeit in a schematic way. In this paper
we formulate BR scaling of \cite{BR91} in a more rigorous way and
offer a scheme that could shed light on the $intrinsic$ density
dependence needed in the HLS/VM theory.

The basic idea in our approach is as follows. We consider a
skyrmion-type Lagrangian with spontaneously broken chiral symmetry
and scale symmetry associated, respectively, with nearly massless
quarks and trace anomaly of QCD. Such a theory may be considered
as an $N_c\rightarrow \infty$ approximation to QCD. Suppose that
the Lagrangian can describe not only the lowest-excitation, i.e.,
pionic, sector but also the baryonic sector and massive vector
meson sector all lying below the chiral scale $\Lambda_\chi$. The
skyrmion description gives not only the single baryon spectra but
also multi-baryon systems including infinite nuclear matter. We
are interested in how low-energy degrees of freedom in many-body
systems behave in dense matter, in particular as the density
reaches a density at which QCD predicts a phase transition from
the broken chiral symmetry to the unbroken chiral symmetry or
chiral restoration in short. For this purpose, one first looks for
the ground state of the many-baryon system in question as a
soliton solution of the Lagrangian and then looks at the
fluctuation of effective fields in various channels of low-energy
excitations with the effect of the background taken into account
self-consistently. As one varies the density of the system, the
parameters of the theory involved in the process must adapt to the
density of the skyrmion background. The problem we are interested
in is whether we can extract from the unified scheme an
information on the intrinsic dependence inherent in effective
field theories matched to QCD , i.e., HLS/VM theory of Harada and
Yamawaki. The hope is that by looking at the structure at the
phase transition and tuning the parameters such that the features
predicted by HLS/VM are reproduced, we can ultimately learn about
the intrinsic dependence of the parameters and hence ``derive" BR
scaling. We will see that some progress in this direction can be
made.

The content of this paper is as follows. In Section 2, we give our
model Lagrangian which is the simplest form of skyrmion
Lagrangian, namely, the original Skyrme Lagrangian implemented
with trace anomaly of QCD. In Section 3, a single skyrmion is
analyzed to define the parameters of the theory at zero density.
The chiral phase transition from an FCC crystal to a half-skyrmion
CC crystal is described in terms of the model Lagrangian in
Section 4. Section 5 describes how the pseudo-Goldstone pion and
the scalar of the trace anomaly (``dilaton") behave in the
skyrmion matter as a function of density. In Section 6, the chiral
restoration transition is made from the inhomogeneous phase -- to
which the FCC crystal collapses -- to the half-skyrmion crystal
phase. The contact with BR scaling, and indirectly with the
parametric property of HLS/VM, is made in this section. Some
concluding remarks are given in Section 7.

\section{The Model Lagrangian}
The starting point of our work is the skyrmion Lagrangian
introduced by Ellis and Lanik~\cite{EL85} and employed by Brown
and Rho~\cite{BR91} for nuclear physics that incorporates the
trace anomaly of QCD. Here we will study the same Lagrangian from
a more modern perspective.

The classical QCD action of scale dimension 4 in the chiral limit
is scale-invariant under the scale transformation
\begin{equation}
x \rightarrow {}^\lambda x = \lambda^{-1} x, \ \ \ \lambda \geq 0,
\end{equation}
under which the quark field and the gluon fields transform with
the scale dimension 3/2 and 1, respectively. The quark mass term
of scale dimension 3 breaks this scale invariance. At the quantum
level, scale invariance is also broken by dimensional
transmutation even for massless quarks, as signaled by a
non-vanishing trace of the energy-momentum tensor. Equivalently,
this phenomenon  can be formulated by the non-vanishing divergence
of the dilatation current $D_\mu$, the so called trace anomaly,
\begin{equation}
\partial^\mu D_\mu = \theta^\mu_\mu
= \sum_q m_q \bar{q}q - \frac{\beta(g)}{g} \mbox{Tr}G_{\mu\nu} G^{\mu\nu},
\label{TraceAnomaly}
\end{equation}
where $\beta(g)$ is the beta function of QCD.

We will implement broken scale invariance into  large $N_c$
physics by modifying the skyrmion Lagrangian,
\begin{equation}
{\cal L} = \frac{f_\pi^2}{4} \mbox{Tr} (\partial_\mu U^\dagger
\partial^\mu U) +\frac{1}{32e^2} \mbox{Tr} [U^\dagger \partial_\mu
U, U^\dagger \partial_\nu U]^2 +\frac{f_\pi^2 m_\pi^2}{4}
\mbox{Tr} (U+U^\dagger-2).
\end{equation}
The chiral field $U=\exp(i\vec{\tau}\cdot\vec{\pi}/f_\pi)$ has
scale dimension 0 and  therefore the Lagrangian respects neither
the scale invariance of QCD nor its breaking
eq.~(\ref{TraceAnomaly}). The current algebra term with two
derivatives is of dimesion 2 and the meson mass term is of
dimension 0, while the Skyrme term with four derivatives has scale
dimension 4.

In order to make the Skyrme model well behaved under the scaling
properties of QCD , we introduce an additional degree of freedom
in the form of a scalar field $\chi$ with a scale dimension 1,
whose coupling to the $U$ fields is defined by,
\begin{eqnarray}
{\cal L}
&=& \frac{f_\pi^2}{4} \left(\frac{\chi}{f_\chi}\right)^2
{\rm Tr} (\partial_\mu U^\dagger \partial^\mu U)
+\frac{1}{32e^2} {\rm Tr}
([U^\dagger\partial_\mu U, U^\dagger\partial_\nu U])^2
+\frac{f_\pi^2m_\pi^2}{4} \left(\frac{\chi}{f_\chi}\right)^3
{\rm Tr} (U+U^\dagger-2)
\nonumber \\
&& \hspace{2cm} +\frac{1}{2}\partial_\mu \chi\partial^\mu \chi
-\frac{1}{4}\frac{m_\chi^2}{f_\chi^2}\bigg[\chi^4\bigg(\ln(\chi/f_\chi)
-\frac{1}{4}\bigg)+\frac{1}{4}\bigg].
 \label{lag-chi}
\end{eqnarray}
We have denoted the non vanishing vacuum expectation value of
$\chi$ as $f_\chi$, a constant which describes, as we shall see,
the decay of the scalar into pions. The second term of the trace
anomaly (\ref{TraceAnomaly}) can be reproduced by the potential
energy $V(\chi)$ for the scalar field, which is adjusted in the
Lagrangian~(\ref{lag-chi}) so that $V=dV/d\chi=0$ and $d^2
V/d\chi^2=m_\chi^2$ at $\chi=f_\chi$.

The vacuum state of the Lagrangian at zero baryon number density
is defined by $U=1$ and $\chi= f_\chi$.  We will take the latter
to be positive and therefore the field values around the vacuum
will be positive. The fluctuations of the pion and the scalar
fields about this vacuum, defined through
\begin{equation}
U=\exp(i\vec{\tau}\cdot\vec{\phi}/f_\pi),
\mbox{\ \ and \ \ }
\chi = f_\chi + \tilde{\chi}
\end{equation}
give physical meaning to the model parameters: $f_\pi$ as the pion
decay constant, $m_\pi$ as the pion mass, $f_\chi$ as the scalar
decay constant, and $m_\chi$ as the scalar mass. For the pions, we
use their empirical values as $f_\pi=93$MeV and $m_\pi=140$MeV. We
fix the Skyrme parameter $e$ to 4.75 from the axial-vector
coupling constant $g_A$ as in ref.~\cite{BJRV84}. However, for the
scalar field $\chi$, no experimental values for the corresponding
parameters are available yet.

The scalar field may be interpreted as a bound state of gluons,
the so-called glueball~\cite{EL85}. If one assumes that it is a
pure glueball, then one can use the gluon condensate $G_0=\langle
0 | (\alpha_s/\pi) G_{\mu\nu} G^{\mu\nu} |0 \rangle$ to restrict
the product of its mass $m_\chi$ by the vacuum expectation value
$f_\chi$ as
\begin{equation}
\textstyle \frac12 f_\chi m_\chi = (\frac{9}{8} G_0)^\frac12.
\label{G0}
\end{equation}

Fluctuating around the vacuum at zero density, the Lagrangian
yields the relevant interactions for the process $\chi \rightarrow
\pi\pi$. In the chiral limit, the coupling from the current
algebra term is given by
\begin{equation}
{\cal L}_{\chi \pi \pi} = \frac{\tilde{\chi}}{f_\chi} \sum_{a=1}^3
(\partial_\mu \phi_a)^2. \label{cpp0}
 \end{equation}
This yields the decay width $\Gamma(\chi \rightarrow \pi\pi)$ as
\begin{equation}
\Gamma(\chi \rightarrow \pi\pi) = \frac{3m_\chi^3}{32 \pi
f_\chi^2} = \frac{m_\chi^5}{48\pi} \frac{1}{G_0}
\label{Gamma0}\end{equation} in the $\chi$ rest frame, where we
used eq.~(\ref{G0}) to eliminate $f_\chi$ in favor of $G_0$.
Introducing  the pion mass into the calculation modifies the
expression slightly \cite{EL85}. Using the ITEP value $G_0 =
0.012$GeV$^4$~\cite{ITEP}\footnote{The more recent ITEP
value~\cite{ITEP2} is $G_0=0.011\pm 0.009$ GeV$^4$ which is
consistent with the older value.}, one obtains
\begin{equation}
\Gamma(\chi \rightarrow \pi\pi)
\approx 0.6 \mbox{GeV} \times (m_\chi/1 \mbox{GeV})^5.
\end{equation}
This equation provides a restriction to the range of the possible
gluonium masses
\begin{equation}
\begin{array}{lcl}
\Gamma(\chi \rightarrow \pi\pi) < 1\mbox{ MeV}
& \mbox{for}
& m_\chi < 0.3 \mbox{ GeV} \\
\Gamma(\chi \rightarrow \pi\pi) > 0.5 m_\chi
& \mbox{for}
& m_\chi > 1 \mbox{ GeV}.
\end{array}
\end{equation}
We see that a light scalar gluonium with the mass below 400 MeV
cannot be detected because of its too narrow decay width, while a
heavy one with the mass greater than 1 GeV cannot be recognized as
a well-defined resonance.

In the following development, we will find that the scalar $\chi$
in this picture cannot describe pure gluonium; it must contain a
light scalar quarkonium component. This ``soft" component cannot
vanish in the chiral broken phase in order to satisfy the symmetry
properties of the fundamental theory \cite{FTS95}. The low-energy
sum rules\cite{NSVZ80}, related to the gluon condensates, are
controlled by the gluonium component and can be used to fit the
parameters of the theory in the free case, i.e. $m_\chi$ and
$f_\chi$ from $G_0$. There can however be a subtle ``mended
symmetry" that puts the mass of $\chi$ nearly degenerate with that
of the $\rho$ meson, i.e., $\sim 700$ MeV~\cite{svec}. It is not
clear what happens in dense medium. Our conjecture~\cite{BR96} is
that as the density is increased, the two components ``decouple"
with only the quarkonium component interacting with matter. We can
think of this as the ``hard" gluonium component being integrated
out in an effective field theoretic sense with their properties
absorbed into the definition of the fields and the parameters
\cite{BK94}. This makes sense as the characteristic mass scale
drops compared with the mass scale of the gluonium. Chiral
symmetry restoration implies the vanishing of the ``soft"
component.

There are some indications that the above conjecture makes sense.
In ref.~\cite{FTS95}, the scalar field is incorporated into a
relativistic hadronic model for nuclear matter not only to account
for the anomalous scaling behavior but also to provide the
mid-range nucleon-nucleon attraction. Then, the parameters
$f_\chi$ and $m_\chi$ are adjusted so that the model fits finite
nuclei. One of the parameter sets is $m_\chi=550$ MeV and
$f_\chi=240$ MeV~(Set A). On the other hand, Song {\em et
al.}~\cite{SBMR97} obtain the ``best" values for the parameters of
the effective chiral Lagrangian with the ``soft" scalar fields so
that the results are consistent with the ``Brown-Rho"
scaling~\cite{BR91}; explicitly, $m_\chi=720$ MeV and $f_\chi=240$
MeV~(Set B). These observations are consistent with the notion
that ``mended symmetry" is operative for the scalar field to be
identified with the sigma field of linear sigma model in certain
``dilaton" limit~\cite{BK94}.

For completeness, we consider also a parameter set of $m_\chi=1$
GeV and $f_\chi=240$ MeV~(Set C) corresponding to a mass scale
comparable to that of chiral symmetry $\Lambda_\chi\sim 4\pi
f_\pi$.

The parameter sets defined above for the model Lagrangian are
summarized in Table 1.

\begin{table}
\caption{Parameter sets of the model Lagrangian}
\begin{center}
\begin{tabular}{lccc}
\hline
  & $m_\chi$ & $f_\chi$ & $G_0$ \\
\hline
set A & 550MeV & 240MeV & 0.004 GeV$^4$ \\
set B & 720MeV & 240MeV & 0.007 GeV$^4$ \\
set C & 1000MeV & 240MeV & 0.012 GeV$^4$ \\
\hline
\end{tabular}
\end{center}
\end{table}

\section{Single Skyrmion with the Scalar Field}
The topological baryon number current associated with the homotopy of
the mapping
$U_0(\vec{r}): S^3(R^3-\{\infty\}) \rightarrow S^3\mbox{  of }SU(2)$
is defined as
\begin{equation}
B^\mu = \frac{1}{24 \pi^2}
\varepsilon^{\mu \nu \lambda \rho} \mbox{Tr} (U_0^\dagger \partial_\nu U_0
U_0^\dagger \partial_\lambda U_0 U_0^\dagger \partial_\rho U_0).
\end{equation}
The soliton solution with the baryon number $B=1$ can be found
by generalizing the spherical hedgehog Ansatz of the original Skyrme
model as
\begin{equation}
U_0(\vec{r})=\exp(i\vec{\tau}\cdot\hat{r} F(r)),
\mbox{ and }
\chi_0(\vec{r}) = f_\chi C(r),
\end{equation}
with two radial functions $F(r)$ and $C(r)$.

Then, the mass of the single soliton can be expressed as
\begin{eqnarray}
M_{B=1} &=& 4\pi
 \int^{\infty}_0  r^2 dr\
\left[ \frac{f_\pi^2}{2}C^2 \bigg( F^{\prime 2}+2\frac{\sin^2 F}{x^2}\bigg)
+ \frac{1}{2e^2}\frac{\sin^2 F}{x^2}
\bigg( \frac{\sin^2 F}{x^2} + 2F^{\prime 2} \bigg)  \right.
\nonumber\\
&& \hspace{2cm} \left.
+  f_\pi^2 m_\pi^2 C^3 (1-\cos F)
+ \frac{{f_\chi}^2}{2} \left( C^{\prime 2}
+ \frac{m_\chi^2}{2} \{ (C^4 \ln C
- \textstyle\frac14) + \frac14 \}\right) \right].
\end{eqnarray}
where the prime means the derivation with respect to $r$.
Variations of $M_{B=1}$ with respect to $F(r)$ and $C(r)$ lead to
the following equations of motion
\begin{eqnarray}
&&\bigg(f_\pi^2 r^2 C^2+\frac{2\sin^2 F}{e^2}\bigg)F^{\prime\prime}(r)
+2f_\pi^2(\ r C^2+r^2CC^{\prime}\ )F^{\prime}(r)
-\frac{2\sin F\cdot \cos F}{e^2}F^{\prime 2}(r)\nonumber\\
&&\hspace{1.5cm}-2f_\pi^2 C^2 \sin F\cdot\cos F
-\frac{2\sin^3 F\cdot\cos F}{e^2r^2}-f_\pi^2m_\pi^2C^3r^2\sin F=0,
\label{F}\end{eqnarray}
for $F(r)$ and
\begin{eqnarray}
&&f_\chi^2 C^{\prime\prime}(r)
+ \frac{2f_\chi^2}{r}C^{\prime}(r)
-f_\pi^2C(r)\bigg(F^{\prime 2}(r)+\frac{2\sin^2 F}{r^2}\bigg)
-3f_\pi^2m_\pi^2(1-\cos F)C^2(r)\nonumber\\
&&\hspace{4cm}+f_\chi^2m_\chi^2C^3(r)\ln C(r)=0,
\label{C}
\end{eqnarray}
for $C(r)$.

At infinity, the fields $U_0(\vec{r})$ and $\chi_0(\vec{r})$
should reach their vacuum values. The asymptotic behavior of the
equations of motion reveals that the radial functions reach the
corresponding vacuum values as
\begin{equation}
F(x) \sim \frac{e^{-m_\pi r}}{r},\  \mbox{ and }\ C(x) \sim
1-\frac{e^{-m_\chi r}}{r}. \label{at_infty}\end{equation}
 In order for the solution to carry a baryon number, $U_0$ has the
value $-1$ at the origin, that is,  $F(x=0)=\pi$, while there is no
such topological constraint for $C(x=0)$. All that is required is
that it be a positive number below 1. The equations of motion tell
us that for small $x$,
\begin{equation}\begin{array}{l}
F(x) \sim \pi - \alpha x + \gamma x^3 + O(x^5), \\
C(x) \sim C_0 + \beta x^2 + O(x^4).
\end{array}
\label{at_origin}\end{equation}

The coupled equations of motion for $F(x)$ and $C(x)$ together
with the boundary conditions (\ref{at_infty}) and
(\ref{at_origin}) can be solved in various different ways. For
example, we can use an iteration method. We first start with
$C_{i=0}(x)=1$ for all $x$ and solve the equation of motion
(\ref{F}) with $C(x)$  fixed as $C_0(x)$ to obtain $F_{i=0}(r)$,
which is nothing but the profile function of the single skyrmion
of the original Lagrangian. Then, we take this $F_{0}(x)$ as
$F(x)$ in the equation of motion (\ref{C}) to obtain $C_{i=1}(x)$.
We repeat this iteration until $F_{i+1}(x)$ and $C_{i+1}(x)$
converge to $F_{i}(x)$ and $C_{i}(x)$. This iteration converges
quite fast and after ten iterations the solution can be found with
a sufficient accuracy.

\begin{figure}
 \centerline{\epsfig{file=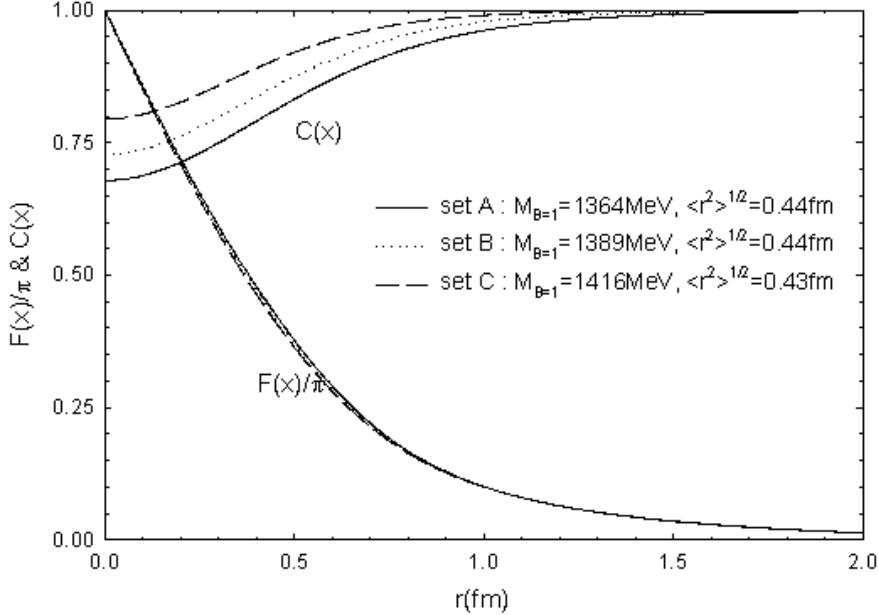,width=12cm,angle=0}}
\caption{Profile functions $F(x)$ and $C(x)$ as a function of
$x$.}
 \label{prof}
\end{figure}

Shown in Fig.~\ref{prof} are the profile functions $F(x)$ and $C(x)$ as
a function of $x(=ef_\pi r)$.
$F(r)$ and consequently the root mean square radius of the
baryon charge
\begin{equation}
\langle r^2\rangle^{1/2} = \left( \int d^3r r^2 B^0(\vec{r}) \right)^{1/2}
\end{equation}
show little dependence on $m_\chi$. On the other hand, the changes in
$C(r)$ and the soliton mass are recognizable. The larger the
scalar mass is, the smaller its coupling to the pionic field and
the less its effect on the single skyrmion. In the limit of
$m_\chi \rightarrow \infty$, the scalar field is completely
decoupled from the pions and the model returns back to the
original one, where $C(r)=1$, $M_{sol}=1479$ MeV and $\langle
r^2\rangle^{1/2}=0.43$ fm. The strong dependence of $C(r)$ on
$m_\chi$ may come from the asymptotic behavior (\ref{at_infty}) at
large $x$. As for the solion mass, note that $M_{sol}$ scales
approximately as $(f_\pi/e)$ and $C^2(r)$ is multiplied to
$f_\pi^2$ in the current algebra term of the Lagrangian. Thus,
$C(r)\leq 1$ reduces the {\em effective} $f_\pi$ inside the single
skyrmion so that the soliton mass decreases accordingly. For
example, for the parameter set $A$, the soliton mass gets 7\%
reduction and we can imagine that the effective $f_\pi$ is reduced
by the same amount in average in the region where the single
skyrmion is located.

\section{FCC Skyrmion Crystal with the Scalar Field}
Consider a crystal configuration made up of skyrmions, where each
FCC lattice site is occupied by a single skyrmion center with
$U_0=-1$ and each nearest neighboring pair is relatively rotated
in isospin space by $\pi$ with respect to the line joining them.
At low density, the system maintains the original configuration,
i.e., it appears as an FCC crystal with the dense centers of
single skyrmions on each lattice sites. At high density, however,
the system undergoes a phase transition to a CC crystal made up of
half-skyrmions. Here, only half of the baryon number carried by
the single skyrmion is concentrated at the original FCC sites,
while the other half is concentrated at the center of the links
connecting these points.

Let us denote the field configuration for the skyrmion field by
$U_0(\vec{x})=\sigma+i\vec{\tau}\cdot\vec{\pi}$ (with $\sigma^2 +
\pi^2=1$)  and $\chi_0(\vec{x})$ for the scalar field. The FCC
configuration we are considering has the symmetries listed in
Tab.~2. There, $2L$ is the size of the single FCC unit cell that
contains 4 skyrmions. Thus, the baryon number density is
$\rho=1/2L^3$. Normal nuclear matter density occurs at $\rho_0
=0.17/$fm$^3$which corresponds to $L\sim 1.43$ fm. Note that the
chiral $\sigma$ field has exactly the same symmetries as those of
the $\chi$.

\begin{table}
\begin{center}
\caption{Symmetries of the FCC skyrmion crystal}
\begin{tabular}{cccc}
\hline
symmetry & space & $U_0$ & $\chi$ \\
\hline
reflection
& $\vec{x} \rightarrow (-x,y,z)$
& $U_0 \rightarrow (\sigma, -\pi_1, \pi_2, \pi_3)$
& $\chi_0 \rightarrow \chi_0$ \\
3-fold axis rotation
& $\vec{x} \rightarrow (y,z,x)$
& $U_0 \rightarrow (\sigma, \pi_2, \pi_3, \pi_1)$
& $\chi_0 \rightarrow \chi_0$ \\
4-fold axis rotation
& $\vec{x} \rightarrow (x,z,-y)$
& $U_0 \rightarrow (\sigma, \pi_1, \pi_3, -\pi_2)$
& $\chi_0 \rightarrow \chi_0$ \\
translation
& $\vec{x} \rightarrow (x+L,y+L,z)$
& $U_0 \rightarrow (\sigma, -\pi_1, -\pi_2, \pi_3)$
& $\chi_0 \rightarrow \chi_0$ \\
\hline
\end{tabular}
\end{center}
\end{table}

As for the constrained fields $(\sigma, \pi_a)$, it is convenient
to work with ``{\em unnormalized}" fields $(\bar{\sigma},
\bar{\pi}_a)$, which can then be ``{\em normalized}" as
\begin{equation}
\sigma = \frac{\bar{\sigma}}{\sqrt{\bar{\sigma}^2 + \bar{\pi}_1^2
+ \bar{\pi}_2^2 + \bar{\pi}_3^2 }}, \label{norm}\end{equation} and
similarly for $\pi_a~(a=1,2,3)$. The field configurations obeying
the above symmetries can be easily found by expanding the
unnormalized fields in terms of Fourier series as~\cite{KS89}
\begin{equation}
\bar{\sigma}(\vec{x}) = \sum_{a,b,c} {\beta}_{abc} \cos(a\pi x/L)
\cos(b\pi y/L) \cos(c\pi z /L),
\label{sigma_bar}\end{equation}
and
\begin{eqnarray}
\bar{\pi}_1(\vec{x}) &=& \sum_{h,k,l} {\alpha}_{hkl} \sin(h\pi x/L)
\cos(k\pi y/L) \cos(l\pi z/L),
\label{pi1} \\
\bar{\pi}_2(\vec{x}) &=& \sum_{h,k,l} {\alpha}_{hkl} \cos(l\pi x/L)
\sin(h\pi y/L) \cos(k\pi z/L),
\label{pi2} \\
\bar{\pi}_3(\vec{x}) &=& \sum_{h,k,l} {\alpha}_{hkl} \cos(k\pi x/L)
\cos(l\pi y/L) \sin(h\pi z/L),
\label{pi3}
\end{eqnarray}
and finally
\begin{equation}
\chi_0(\vec{x}) = \sum_{a,b,c} {\gamma}_{abc} \cos(a\pi x/L)
\cos(b\pi y/L) \cos(c\pi z /L).
\label{chi}\end{equation}

The symmetries of Tab.~2 restrict the modes appearing in
eqs.~(\ref{sigma_bar}-\ref{chi}) as follows;
\renewcommand{\theenumi}{\arabic{enumi}}
\renewcommand{\labelenumi}{(M\theenumi)}
\begin{enumerate}
\item if $h$ is even, then $k,{}l$ are restricted to odd numbers
and $a,{}b,{}c$ are to even numbers,
\label{M1}
\item if $h$ is odd, then $k,{}l$ are restricted to even numbers
and $a,{}b,{}c$ are to odd numbers.
\label{M2}
\end{enumerate}
Furthermore, ${\alpha}_{hkl}={\alpha}_{hlk}$ and
${\beta}_{abc}={\beta}_{bca}={\beta}_{cab}
={\beta}_{acb}={\beta}_{cba}={\beta}_{bac}$.

We can locate, without loss of generality, the centers of the
skyrmions at the corners of the cube and at the centers of the
faces by letting ${\sigma}=-1$ and ${\pi}_i(i=1,2,3)=0$ at those
points. For the skyrmion field to have a definite integer baryon
numbers per site, we should have ${\sigma}=+1$ and
${\pi}_i(i=1,2,3)=0$ at points such as $(L,0,0)$. This produces
the constraint,
\begin{equation}
\sum_{a,b,c=\mbox{\scriptsize even}} \bar{\beta}_{abc}=0 \ .
\end{equation}
As for  $\gamma_{abc}$ associated with the scalar field,
there is no such constraint, but the coefficients should be
arranged to satisfy $\chi_0 \geq 0$, a consequence of our choice
of vacuum.

 If we had only the modes~(M\ref{M2}) in the expansion, the
configuration would then have an additional symmetry, namely,
under the translation $\vec{x}\rightarrow (x+L,y,z)$ the field
undergoes an $O(4)$ rotation by $\pi$ in the $\sigma, \pi_1$
plane. Furthermore, in order to satisfy the constraint $\chi \geq
0$, $\chi$ must vanish identically. This configuration corresponds
to the {\em half-skyrmion} CC as explained above. Because of this
additional symmetry, physical quantities such as the local baryon
number density and the local energy density become completely
identical around the points with $\sigma=-1$ and the points with
$\sigma=+1$. Thus, one half of the baryon number carried by a
single skyrmion is concentrated at the sites where the centers of
the single skyrmion are expected to be in the FCC crystal. The
other half of the baryon number is now concentrated on the links
connecting those points, where the $\sigma$ takes the value $+1$
and, in the original FCC configuration, the local baryon number
density is rather low. As a consequence, the expectation value
$\langle \sigma \rangle$ goes to zero, signaling the restoration
of the spontaneously  broken chiral symmetry. Both modes, (M1) and
(M2), are included in the actual numerical procedure which we
define next. The half-skyrmion crystal configuration arises at
high density where the expansion coefficients associated with the
modes (M1) become small.

In order to obtain the coefficients we minimize the energy per baryon $E/B$
given by
\begin{eqnarray}
E/B &=& - \frac14 \int_{Box} d^3r {\cal L}_{{\rm M}}(U_0,\chi_0)
\nonumber \\
&=& \frac{1}{4}
\int_{Box} d^3x \left\{
\frac{f^2_\pi}{4} \left( \frac{\chi_0}{f_\chi} \right)^2
\mbox{Tr}(\partial_i U_0^\dagger  \partial_i U_0)
+ \frac{1}{32e^2} \mbox{Tr} \left[ U_0^\dagger \partial_i U_0,
 U_0^\dagger \partial_j U_0 \right]^2
\right.
\nonumber \\
&& \left.
\hskip 5em +\frac{f_\pi^2 m_\pi^2}{4}
\left(\frac{\chi_0}{f_\chi} \right)^3 \mbox{Tr} (2-U^\dagger_0 - U_0)
+ \frac12 \partial_i \chi_0 \partial_i \chi_0 + V(\chi_0) \right\}
\label{EoverB}\end{eqnarray}
by taking the coefficients  of the expansions
as variational parameters. In eq.~(\ref{EoverB}), the subscript
`box' denotes that the integration is over a single FCC box and
the factor 1/4 in front appears because the box contains
baryon number four. We employ ``the down-hill simplex
method"~\cite{NR} for the minimization process.

\begin{figure}
\centerline{\epsfig{file=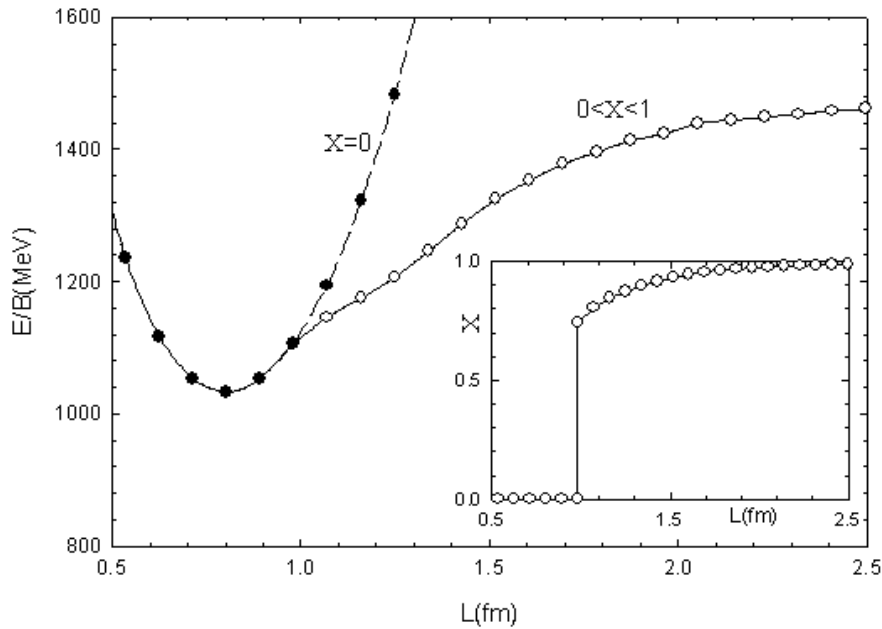,width=12cm,angle=0}} \
\caption{Energy per single skyrmion as a function of $L$. The
results are obtained by minimizing $E/B(X,L)$ with respect to $X$
for a given $L$. }
 \label{Toy-2}
 \end{figure}

{Before going into further details, let us do a rough study of the
role the chiral field plays in the phase structure of the system.
For this, we take $\chi_0$ to be a constant
\begin{equation}
\chi_0/f_\chi = X.
\end{equation}
Then eq.(\ref{EoverB}) can be approximated to
\begin{equation}
E/B(X,L) =  X^2 (E_2/B) + (E_4/B) + X^3 (E_m/B) +
(2L^3) \left(X^4 (\mbox{ln}X - \textstyle \frac14) + \frac14\right),
\end{equation}
where $E_2$, $E_4$ and $E_m$ are, respectively, the contributions
from the current algebra term, the Skyrme term and the pion mass
term of the Lagrangian to the energy of the skyrmion system and
$(2L^3)$ is the volume occupied by a single skyrmion. It can be
understood as an effective potential for the chiral field {\em in
medium}, modified by the coupling of the scalar to the background
matter. With the parameter values obtained in ref.\cite{LPMRV03}
for the Skyrme model without the scalar fields, the effective
potentials $E/B(X)$ behave as shown in Fig. \ref{Toy-1}. At low
density (larger $L$), the minimum of the effective potential is
located slightly from $X=1$. As the density increases,  the
effective potential $V(\chi)$ develops another minimum at $X=0$
which was an unstable extremum of the potential in free space. At
$L\sim 1$ fm, the newly developed minimum can compete with the one
near $X \sim 1$. At higher density, the minimum gets shifted to
$X=0$ where the system gets stabilized.

\begin{figure}
\centerline{\epsfig{file=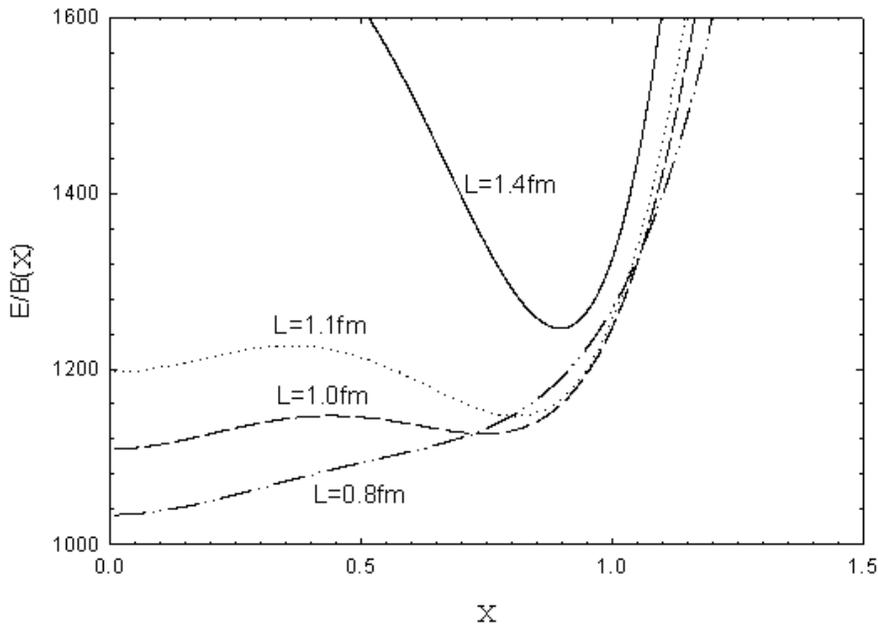,width=12cm,angle=0}} \
\caption{Energy per single skyrmion as a function of the scalar
field $X$ for a given $L$. The results are obtained with the
$(E_2/B)$, $(E_4/B)$, and $(E_m/B)$ of ref.\cite{LPMRV03} and with
the parameter sets B.}
 \label{Toy-1}
 \end{figure}

In Fig. \ref{Toy-2}, we plot $E/B(X_{min},L)$ as a function of
$L$. The figure in a small box is the corresponding value of
$X_{min}$ as function of $L$. There we see an explicit
manifestation of a first-order phase transition. Although the
discussion is perhaps a bit too naive, it essentially encodes the
same physics as in the more rigorous treatment of $\chi_0$ given
below. }

\begin{figure}
\centerline{\epsfig{file=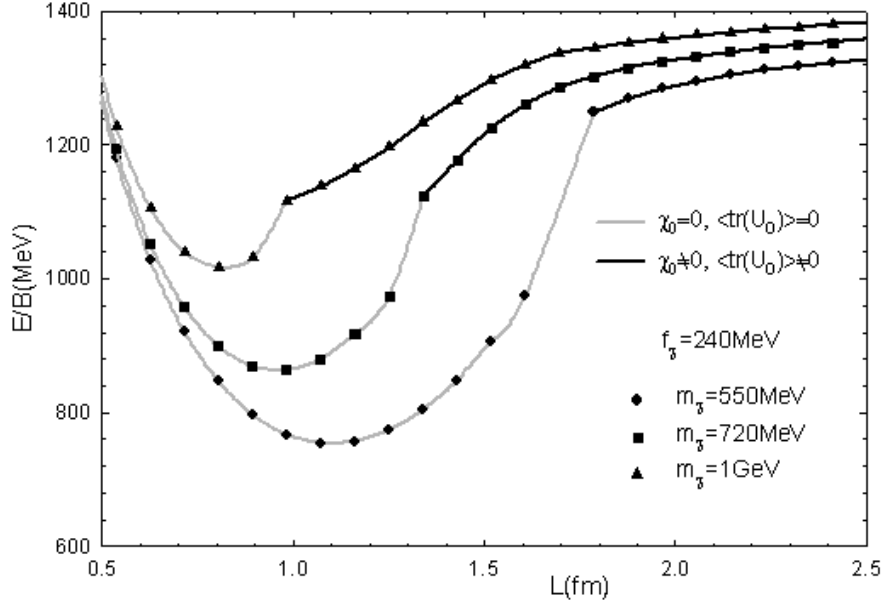,width=12cm,angle=0}} \
\caption{Energy per single skyrmion as a function of the size
parameter $L$. The solid circles, solid squares, solid triangles
are the results obtained with the parameter sets A, B, and C,
respectively.} \label{EperB}
 \end{figure}

Given in Fig.~\ref{EperB} is the energy per baryon, $E/B$, as a
function of the FCC box size parameter $L$. Each point corresponds
to the lowest energy crystal configuration for the given value of
$L$. The solid circles, solid squares and solid triangles are
obtained with the parameter sets A, B, and C, respectively.
Furthermore, the black solid lines correspond to the single
skyrmion FCC phase, while the gray lines to the half-skyrmion CC
phase. As we squeeze the system from $L=6$ on, the skyrmion system
undergoes at $L=L_{pt}$ a phase transition from the FCC single
skyrmion configuration to the CC half-skyrmion configuration. The
transition appears to be first order. In the half-skyrmion phase,
$\chi_0(\vec{r})$ vanishes. The energy of the
system comes only from the Skyrme term and the scalar field
potential. The former roughly scales as $\sim 1/L$ and the latter
exactly as $L^3$ (the volume of the box); explicitly,
\begin{equation}
E/B \sim a/L + 2V(\chi_0=0) L^3,
\end{equation}
$a$ being constant. After the phase transition, the energy per
baryon $E/B$ continues to decrease~(even faster) and reaches its
minimum point at $L=L_{min}$. The precise values for the phase
transition point $L_{pt}$ and the minimum energy point $L_{min}$
depend on the parameters and are therefore different for the various
sets.

The incorporation of the scalar field into the Skyrme model
Lagrangian produces quite dramatic effects on the properties of
skyrmion matter. For example, the energy per baryon drops down to
$\sim 700$ MeV. This can be understood since the potential energy
between the skyrmions amounts to nearly 50\% of the skyrmion mass
and comes from the medium-range attraction generated by the scalar
field.

\begin{figure}
\centerline{\epsfig{file=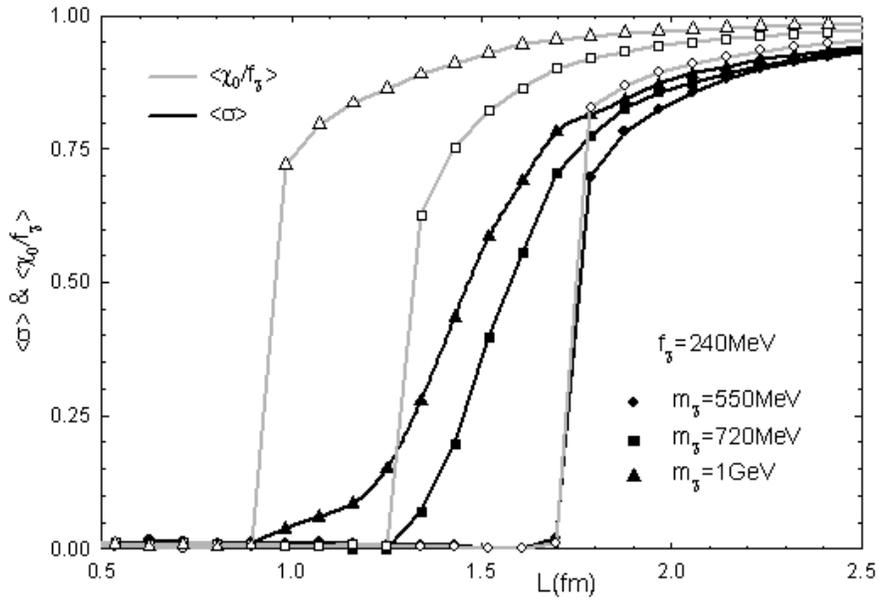,width=12cm,angle=0}} \
\caption{$\langle \sigma\rangle$ and $\frac{\chi_0}{f_\chi}$ as a
function of the size parameter $L$. } \label{sigma}
 \end{figure}

From here on we denote by $\langle Q \rangle$ the average value of a quantity
$Q(\vec{r})$ over the FCC box defined by
\begin{equation}
\langle Q \rangle = \frac{1}{8L^3} \int_{Box} d^3 r Q(\vec{r}).
\end{equation}
In Fig.~\ref{sigma}, we represent $\langle \sigma \rangle$ and
$\langle\chi_0/f_\chi\rangle$ -- the averaged values of
$\sigma(\vec{r})$ and $\chi_0(\vec{r})$ over space -- as a
function of $L$.  Both quantities go to zero at the {\em same}
phase transition point $L=L_{pt}$. However, for larger values of
$m_\chi$, the transition properties of the two quantities look
different. Note that while $\sigma(\vec{r})\neq 0$ locally and
$\langle \sigma \rangle=0$ is obtained by the averaging process,
$\langle \chi_0/f_\chi \rangle=0$ results because
$\chi_0(\vec{r})=0$ throughout the whole space. We will take
$\langle \chi_0 \rangle$ as the {\em effective} value of $f_\chi$,
i.e., $f_\chi^*$. We interpret this result as the vanishing of the
``soft" part of the gluon condensate tied to the quarkonium
component.

\section{Properties of Pions and Scalar in Skyrmion Matter}

In this section, we study the properties of the pionic and scalar
excitations in dense matter. For this purpose, we incorporate the
fluctuations on top of the static  skyrmion
crystal as in \cite{LPMRV03}. This can be achieved using the Ansatz
\begin{equation}
U(x) = \sqrt{U_\pi(x)} U_0(\vec{x}) \sqrt{U_\pi(x)}, \ \ \ \ \
\chi(x) = \chi_0(\vec{x})+\tilde{\chi}(x),
\end{equation}
where $U_\pi=\exp(i\vec{\tau}\cdot\vec{\phi}/f_\pi)$ as in
ref.~\cite{LPMRV03}.
Hereafter, we will drop the tilde to denote the $\chi$ fluctuation.

Expanding the Lagrangian~(\ref{lag-chi}) up to the second order in
the fluctuating fields, we obtain
\begin{equation}
{\cal L}(U, \chi)={\cal L}_{{\rm M}}(U_0, \chi_0)+{\cal L}_{{\rm
M}, \chi} +{\cal L}_{{\rm M}, \phi}+{\cal L}_{{\rm M}, \phi\chi},
\label{Lpi}
\end{equation}
where the subscript `M' denotes the matter field, ${\cal
L}_{\rm M}$ the Lagrangian density for the static configuration
for the background matter and the various terms are given by
\begin{eqnarray}
\nonumber\\
{\cal L}_{{\rm M}, \phi}
&=&
\frac{1}{2} G_{ab} \dot{\phi}_a \dot{\phi}_b
 - \frac{1}{2} H^{ij}_{ab}
     (\vec{x}) \partial_i \phi_a \partial_j \phi_b
 - \frac{1}{2} \bigg( \frac{\chi_0}{f_\chi} \bigg)^3
      m_\pi^2 \sigma(\vec{x}) \phi_a^2
\nonumber\\
&& \hskip 7em + \frac{1}{2f_\pi^2}
 \epsilon_{abc} \partial_\mu \phi_a \phi_b V^{\mu}_c(\vec{x})
\\
{\cal L}_{{\rm M},\chi}
&=& \frac{1}{2} \partial_\mu \chi \partial^\mu \chi
-\frac{1}{2} \bigg[ m_\chi^2
 \bigg(\frac{\chi_0}{f_\chi}\bigg)^2
 \bigg(1+3\ln(\chi_0/f_\chi)\bigg)
 +\frac{2}{f_\chi^2} \frac{f_\pi^2}{4} {\rm Tr}
(\partial_i U^\dagger_0 \partial_i U_0)
\nonumber\\ && \hskip 10em
 +\frac{6\chi_0}{f_\chi^3} f_\pi^2m_\pi^2 (1-\sigma(\vec{x}))
  \bigg] \chi^2
\\
{\cal L}_{{\rm M}, \chi\phi}
&=& \bigg( \frac{2\chi_0}{f_\chi^2}
 \frac{f_\pi}{4} {\rm Tr}(i(L_i-R_i) \tau^a) \bigg)
   \chi \partial^i \phi_a
+ \bigg( \frac{3\chi_0^2}{f_\chi^3} \frac{f_\pi m_\pi^2}{2}
   \mbox{Tr}(i\tau_a U_0) \bigg) \chi\phi^a.
\end{eqnarray}
Here,
\begin{eqnarray}
G_{ab}(\vec{x}) &=& \left( \frac{\chi_0}{f_\chi} \right)^2
( \delta_{ab} + g_{ab}(\vec{x}) )
\displaystyle + \frac{1}{32e^2f_\pi^2}{\rm Tr}
( [R_i,\tilde{\tau}^a] [R_i,\tilde{\tau}^b] ),
\label{G} \\
H^{ab}_{ij}(\vec{x}) &=&  G^{ab}\delta_{ij}
+ \frac{1}{32e^2f_\pi^2}{\rm Tr}\bigg(
[R_i , R_j][\tilde{\tau}^a, \tilde{\tau}^b]
-[R_i, \tilde{\tau}^b][R_j,\tilde{\tau}^a]\bigg),
\label{H} \\
V^a_i(\vec{x}) &=&\bigg(\frac{\chi_0}{f_\chi}\bigg)^2
\frac{i}{4}f_\pi^2 {\rm Tr}[(L_i + R_i )\tau^a]
 + \frac{i}{16e^2} {\rm Tr} \bigg([L_j, \tau^a][L_i, L_j]
 + [R_j, \tau^a][R_i, R_j] \bigg),
\label{V}
\end{eqnarray}
where for later convenience, we have defined a part of $G^{ab}$
separately as
\begin{equation}
g_{ab}(\vec{x}) = \textstyle \frac{1}{4} {\rm Tr} ( \tau_a U_0
\tau_b U_0^\dagger-\tau_a \tau_b ) = -(\pi^2 \delta_{ab} - \pi_a
\pi_b)
\end{equation}
which comes from the current algebra term in the Lagrangian. Here
$L_i$ and $R_i$ are defined by
\begin{equation}
L_i=(\partial_i U_0^\dagger) U_0, \ \ \ \ R_i=(\partial_i U_0)
U_0^\dagger, \label{defLR}
\end{equation}
in terms of the background matter fields.

The previous equations show how the medium, represented by the
background skyrmion field, influences the elementary excitations.
In order to get an idea of the expected results we show the tree
level approximation where we have substituted the background field
by its space average over the cell. In this way we obtain closed
form expressions for the {\em  in-medium parameters} showing their
relation to their vacuum values.

Due to the symmetry of the background field, $\langle
\mbox{Tr}(i\tau_a U_0) \rangle$ vanishes in the averaging
procedure and therefore there are no terms  with $\chi\phi_a$ in
${\cal L}_{M,\chi\phi}$ to this order. The relevant terms arise
from ${\cal L}_{M,\phi}$ and ${\cal L}_{M,\chi}$ and can be
written as

\begin{eqnarray}
{\cal L}_{M,\phi} &=&
\frac{Z_\pi^2}{2} \dot{\phi}_a \dot{\phi}_a
- \frac{m^{*2}_\pi Z_\pi^2}{2} \phi^2_a \cdots,
\\
{\cal L}_{M,\chi} &=&
\frac{1}{2} \dot{\chi} \dot{\chi}
- \frac{m^{*2}_\chi}{2} \chi^2 \cdots.
\end{eqnarray}
where the pion wave function renormalization constant $Z_\pi$,
the in-medium pion~ mass $m_\pi^*$ and scalar mass $m_\chi^*$
are defined as
\begin{eqnarray}
Z_\pi^2 &=& \left\langle \left(\frac{\chi_0(\vec{x})}{f_\chi} \right)^2
\textstyle (1 - \frac23 \pi^2(\vec{x})) \right\rangle
\equiv \left( \frac{f_\pi^*}{f_\pi} \right)^2,
\label{f*/f} \\
m_\pi^{*2} Z_\pi^2 &=&
\left\langle \left(\frac{\chi_0(\vec{x})}{f_\chi} \right)^3
\sigma(\vec{x})\ {m_\pi^2} \right\rangle, \label{mp*/mp}\\
m_\chi^{*2} &=& \left\langle
m_\chi^2 \left(\frac{\chi_0(\vec{x})}{f_\chi} \right)^2
\bigg( 1 + 3 \ln (\chi_0(\vec{x})/f_\chi) \bigg)
+\frac{2}{f_\chi^2} \frac{f_\pi^2}{4} {\rm Tr}
(\partial_i U^\dagger_0 \partial_i U_0) \right.
\nonumber\\
&& \hskip 10em \left.
 +\frac{6\chi_0}{f_\chi^3} f_\pi^2m_\pi^2 (1-\sigma))
   \right\rangle\label{mc*/mc}  .
\end{eqnarray}
The wave function renormalization constant $Z_\pi$ gives the ratio
of the in-medium pion decay constant $f_\pi^*$ to the free one, and
the above expression arises from the current algebra term with $f_\pi$
in the Lagrangian. The other two equations reflect how the medium affects
the effective masses of the mesons.

\begin{figure}
\centerline{\epsfig{file=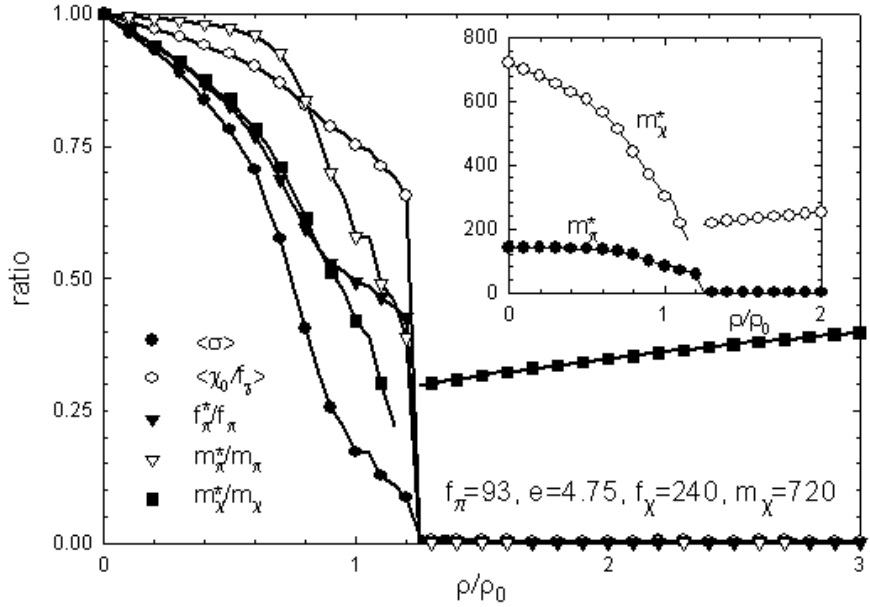,width=12cm,angle=0}}
\caption{The ratios of the in-medium parameters to the free space
parameters. The graph in a small box shows the masses of the pion
and the scalar.} \label{stars}
 \end{figure}

In Fig.~\ref{stars} we show the ratios of the in-medium parameters
relative to their free-space values. Only the results obtained
with the parameter set B are presented. The other parameter sets
yield similar results except that $L_{pt}$ takes different
values~\footnote{Since the parameters are not uniquely given and
will strongly depend on the details of the Lagrangian, one should
not take the precise value of $L_{pt}$ seriously.}. It is
important to note that at $L=L_{pt}$, $all$ the ratios approach
zero except for $m_\chi^*/m_\chi$.

Let us analyze these ratios at the light of the above equations
and related work. As stated before $\chi_0$ and $\langle\sigma\rangle$ 
vanish in the dense matter phase. This is the reason for the vanishing of
two of the above ratios. The non-vanishing of $m_\chi^*/m_\chi$ is
due to the existence of a pure background term which appears in
the contribution to the in-medium scalar mass. This term
substitutes in the dense matter phase the trace anomaly relation
eq.~(\ref{G0}) by
\begin{equation}
\frac{1}{2} m_\chi^*f_\chi = \sqrt{\bigg\langle\frac{f_\pi^2}{8} {\rm Tr}
(\partial_i U^\dagger_0 \partial_i U_0)\bigg\rangle},
\end{equation}
which remains finite. In line with our previous arguments, we see
that quantities related to the condensate in free space become
tied to chiral fields in dense matter.

The extra terms in eq.~(\ref{mc*/mc}), which are not proportional
to $m_\chi^2$, come from the properly scaled current algebra and
the pion mass terms. In the (matter-free) vacuum these terms
describe the couplings of $\chi$ to the $\pi$'s. In dense medium
$\chi$ couples to the static background matter fields which
contribute to the $\chi$ mass.

The vanishing of the pion mass, which already occurred in the pure
Skyrme model calculation due to the vanishing of the 
$\langle\sigma\rangle$, shows the chiral behavior of $\chi_0$, 
since the transition density for both phenomena is the same.

The vanishing of $f^*_\pi/f_\pi$ represents the main qualitative
difference from our previous calculation~\cite{LPMRV03} and
requires a more careful analysis. In our previous calculation the
ratio was found to saturate to $\sim 2/3$ at high density. The
scalar field shifts the effective potential of the pion in matter
from one where the symmetry of the ground state is spontaneously
broken, to one where it is not, i.e., in the language of the sigma
model, the scalar field ``lifts" in the medium the matter-free
vacuum constraint . One may interpret the scalar field 
$\chi_0$ as the ``radius field" of Chanfray {\it et al\/}~\cite{CEG01}
~\footnote{It is easy to see
this in the work of Beane and van Kolck~\cite{BK94} who relate
$\Sigma\propto U\chi$ to the sigma model fields $\sigma
+i\vec{\tau}\cdot\vec{\pi}$ with of course no sigma model
constraint on the radius of the chiral circle. According to this
interpretation, $\chi_0$ is, trivially, what lifts the vacuum
constraint by definition. Furthermore, if we take $\langle
\chi_0/f_\chi\rangle$ as $1+\theta/f_\pi$ of ref.\cite{CEG01}, we
can expand $f^*_\pi/f_\pi$ of Eq.(\ref{f*/f}) as
\begin{equation}
f^*_\pi/f_\pi \sim \left( 1 + \theta/f_\pi - \textstyle\frac13
\vec{\pi}^2 \right).
\end{equation}
This is related to eq.(43) of ref.\cite{CEG01}. (In order for the
comparison, we have to take into account that $\vec{\pi}$
corresponds to their $\vec{\phi}/f_\pi$.) Note also that, in
ref.\cite{CEG01}, the last term proportional to the ``{\em nuclear
virtual pion condensation}" cannot be derived by their ``shifted"
$\theta$; it is put in by hand. }.

The phenomenon discussed above is closely related to ``Brown-Rho"
scaling~\cite{BR91}. In the description of \cite{BR91}, the
density dependence comes solely from the change in the mean field
$\chi^*$ with the corresponding change to the skyrmion structure
ignored. Our present result corrects and gives a precise meaning
to the scaling relation of \cite{BR91}. Similarly eq.~(\ref{f*/f})
compares to the corresponding $f_\pi^*/f_\pi$ of ref.~\cite{BR91}.
Again an additional factor, $(1-\frac23 \vec{\pi}^2 (\vec{x}) )$,
associated with the background property appears as a correction.

At low matter density,  the ratio $f_\pi^*/f_\pi$ can be
fit to a linear function
\begin{equation}
\frac{f_\pi^*}{f_\pi} \sim 1 - 0.24(\rho/\rho_0) + \cdots
\end{equation}
At $\rho=\rho_0$, this yields $f_\pi^*/f_\pi=0.76$ which is to be
compared with $f_\pi^*/f_\pi \approx 0.78$ of ref.~\cite{BR91}.
The ratio $m_\chi^*/m_\chi$ scales similarly to $f_\pi^*/f_\pi$ up
to $\rho \sim \rho_0$. In Tab.~3, we list the slopes of the
ratios. The inset figure in Fig.~\ref{stars} shows the behavior of
the masses $m_\pi^*$ and $m_\chi^*$. They become $nearly$
degenerate close to the ``critical" density. This near degeneracy
may be indicative of the ``mended symmetry" discussed by Beane and
van Kolck~\cite{BK94}. Note however that in contrast to the
dilaton limit of \cite{BK94}, the scalar mass remains finite at
the phase transition as pointed out before while the pion mass
vanishes~\footnote{Our skyrmion Lagrangian is not applicable after
the phase transition, so the spectra in the Wigner phase cannot be
considered physically meaningful.}.

\begin{table}
\caption{The slope of the ratios near the origin}
\begin{center}
\begin{tabular}{ccccc}
\hline
   & $f_\pi^*/f_\pi$ & $m_\pi^*/m_\pi$ & $f_\chi^*/f_\chi$
& $m_\chi^*/m_\chi$ \\
\hline
set A & - 0.21 & - 0.01 & - 0.14 & - 0.25 \\
set B & - 0.25 & - 0.03 & - 0.14 & - 0.28 \\
set C & - 0.12 & - 0.005 & - 0.07& - 0.14 \\
\hline
\end{tabular}
\end{center}
\end{table}

\begin{figure}
 \centerline{\epsfig{file=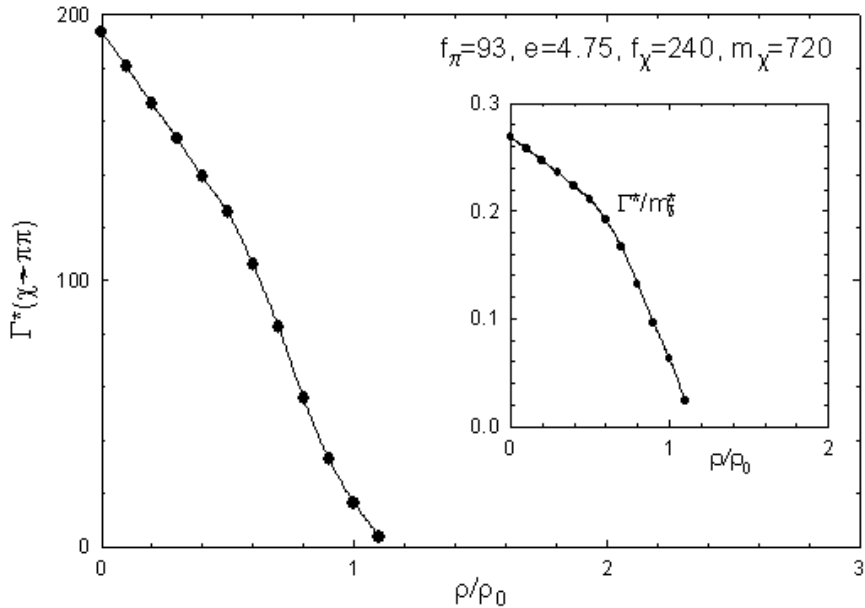,width=12cm,angle=0}}
\caption{The in-medium decay width $\Gamma^*(\chi
\rightarrow\pi\pi)$ as a function of $\rho$. We use
eq.(\ref{exact}) (not eq.(\ref{approx})) to evaluate the
quantities.} \label{Gamma}\end{figure}

It is interesting to look at the decay of $\chi$ into two pions
in the medium. Gathering the terms with a scalar field and two pion
fields, we get the Lagrangian density for the process
$\chi \rightarrow \pi\pi$
\begin{equation}
{\cal L}_{M,\chi\pi^2} = \frac{\chi_0}{f_\chi^2}
(\delta_{ab}+g_{ab}) \chi \partial_\mu \phi_a \partial^\mu \phi_b
+\frac{3\chi_0^2}{2f_\chi^3} m_\pi^2 \sigma(\vec{x}) \chi \phi_a^2,
\end{equation}
To compare this result with eqs.~(\ref{cpp0},\ref{Gamma0}) we take
only the first term, i.e., the current algebra term, in the
Lagrangian. Averaging the space dependence of the background field
configuration modifies the coupling constant by a factor $\langle
(\chi_0/f_\chi)(1+g_{11})\rangle = \langle
(\chi_0/f_\chi)(1-\frac23\pi^2)\rangle $. Taking into account the
appropriate wave function renormalization factors, $Z_\pi$, and
the change in the scalar mass, we obtain, in the chiral limit, an
in-medium decay width of the form
\begin{equation}
\Gamma^* (\chi \rightarrow \pi\pi) =
 \frac{3m_\chi^{*3}}{32\pi f_\chi^2}
 \left| \frac{\la(\chi_0/f_\chi)(1-\frac23\pi^2)\ra}
{\la(\chi_0/f_\chi)^2 (1-\frac23\pi^2)\rangle} \right|^2 .
\label{exact}\end{equation}
 Applying the naive mean field
approximation for averaging, we arrive at
\begin{equation}
\Gamma^*(\chi \rightarrow \pi\pi) \approx \frac{3
m_\chi^{*3}}{32\pi f_\chi^{*2}}. \label{approx}\end{equation} This
expression corresponds to the free one where  $m_\chi$ and
$f_\chi$ are replaced by the in-medium quantities $m_\chi^*$ and
$f_\chi^*$. We show in Fig.~\ref{Gamma} the in-medium decay width
predicted with the parameter set B. In the region $\rho \geq
\rho_{pt}$ where $\chi_0=0$, $\Gamma^*$ cannot be defined to this
order. Near the critical point, the scalar becomes an extremely
narrow-width excitation, a feature which has been discussed in the
literature as a signal for chiral restoration~\cite{HK01,F03}.

\section{From an Inhomogeneous Phase to the Half-Skyrmion Crystal}
Up to this point, we have described  the background field as an
FCC skyrmion crystal. However, as can be seen in Fig.~\ref{EperB},
in the range of $L>L_{min}$, the pressure of the skyrmion system
$-\partial E/\partial V$ is negative, which implies that the
system is unstable. As discussed in ref.~\cite{LPMRV03}, the
system chooses to avoid instability by going into an inhomogeneous
phase instead of remaining in the homogeneous crystal structure.
In this stable phase some part of the volume is occupied by the
matter with a density $\rho_{min}\equiv 1/2L_{min}^3$ and the rest
of the volume is empty. We call this ``inhomogeneous phase." In
this inhomogeneous phase, the spatial average is calculated by
\begin{equation}
\langle Q\rangle^{(i)} = 1-
(\langle Q \rangle^{(h)}_{vac} -  \langle Q \rangle^{(h)}_{min})
(\rho/\rho_{min}) ,
\end{equation}
where the superscripts `(i)' and `(h)' denote that the average
values are evaluated for the inhomogeneous and homogeneous phases,
respectively, and the subscripts `vac' and `min' denote,
respectively, the vacuum and the lowest energy configuration at
$\rho=\rho_{min}$.

What comes out from this naive averaging are the the relevant
quantities in the free space and those in the minimum energy
half-skyrmion phase. Making the averages in
eqs.~(\ref{f*/f}-\ref{mc*/mc}), we obtain the simple scaling
\begin{equation}
(f_\chi^*/f_\chi)^{2(i)} = (f_\pi^*/f_\pi)^{2(i)} =
(m_\pi^* f_\pi^*/f_\pi m_\pi)^{2(i)}=
1- \rho/\rho_{min},
\label{scaling1}\end{equation}
and
\begin{equation}
(m_\chi^*/m_\chi)^{2(i)} = 1 - (1-(m_\chi^*/m_\chi)^{2(h)}_{min})
\rho/\rho_{min}, \label{scaling2}\end{equation} where we have
exploited that only the ratio $(m_\chi^*/m_\chi)_{min}^{2(h)}$ is
non-vanishing.

\begin{figure}
\centerline{\epsfig{file=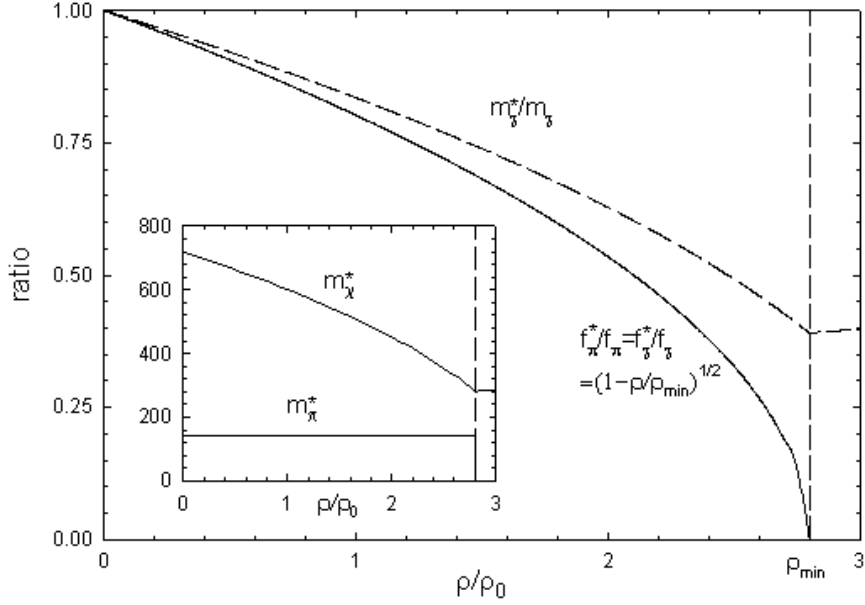,width=12cm,angle=0}}
\caption{The scaling of the in-medium parameters in the
inhomogeneous skyrmion matter.  The graph in a small box shows the
masses of the pion and the scalar.} \label{inhom}\end{figure}

We show in Fig. \ref{inhom} the schematic scaling behavior of
physical quantities in the inhomogeneous background matter. Here,
the phase transition occurs at $\rho=\rho_{min}$ from the
inhomogeneous phase to the half-skyrmion phase. The scaling is
given simply by eq.(\ref{scaling1}) (which depends only on
$\rho_{min})$ and eq.(\ref{scaling2}). Note that in this naive
approximation, the pion mass stays unchanged up to
$\rho=\rho_{min}$ and then it drops to zero.
%
At low density, we have
\begin{eqnarray}
&&\frac{f_\chi^*}{f_\chi}
\approx \frac{f_\pi^*}{f_\pi}
\approx
1 - \frac{1}{2} \frac{1}{2.8} (\rho/\rho_0) +\cdots
= 1 -0.18 (\rho/\rho_0) +\cdots ,
\\
&&\frac{m_\pi^*}{m_\pi} \approx 1,
\\
&& \frac{m_\chi^*}{m_\chi} \approx 1 -0.15(\rho/\rho_0)+\cdots,
\end{eqnarray}
where we have used the parameter set B for which
$\rho_{min}/\rho_0 \approx 2.8$. These results are consistent with
the scaling found by Song {\it el al\/}~\cite{SBMR97}. See
eqs.~(16-18) of ref.~\cite{SBMR97}.

It would be desirable to see how the same quantities scale near
the phase transition at $\rho=\rho_{min}$. Unfortunately, our
naive expression does not allow us to do so.

\section{Further Remarks and Conclusion}

This paper represents a significant step toward a systematic
understanding of in-medium properties of hadrons from the
perspective of a unified theory of the hadronic interactions. In
our previous work ~\cite{LPMRV03}, our starting point was the
Skyrme Lagrangian whose parameters are defined in order to
reproduce mesonic properties in the $B=0$ sector and single
nucleon properties in the $B=1$ sector and then we extended the
description to a skyrmion matter as a way to understand the
properties of nuclear matter and the in-medium properties of
mesons and nucleons. There the phase transition from Goldstone
phase to Wigner phase took place from an FCC crystal state to a
half-skyrmion CC crystal structure. The chiral-symmetry restored
phase supported a non-vanishing pion decay constant. We
interpreted this phase as an analog to the pseudo-gap phase of
high $T_c$ superconductivity. However there may be a different
interpretation. The fact that chiral symmetry is restored with a
non-vanishing pion decay constant is reminiscent of Georgi's
vector limit where chiral restoration occurs with the excitation
of scalar partners of the Goldstone pions with an equal and
non-vanishing decay constant~\cite{georgi}. It is known however
that the Georgi vector limit is not consistent with the chiral
Ward identity~\cite{HY:PR}, so we believe that if the model is
viable, a more likely possibility is the pseudo-gap type
realization.

In this paper, we extended the model studied in \cite{LPMRV03} to
one in which the trace anomaly of QCD is implemented in terms of a
dilaton scalar field. The idea is very similar to what was adopted
in \cite{BR91} and is closely based on the Lagrangian introduced
by Ellis and Lanik in 1985~\cite{EL85}.

The first question we had to address was how to interpret, in the
context of the issue at hand, the scalar interpolating field
introduced to represent the conformal symmetry breaking in QCD in
terms of physical particles. It is generally understood that the
$B=0$ sector is dominated by pions and gluonium with the trace
anomaly more or less saturated by the glueball
contribution~\cite{miransky,NSVZ80}. However chiral symmetry
breaking and scale symmetry breaking must be intricately tied,
requiring the presence of a quarkonium component. How the two
components are mixed is presently a hotly debated issue and is not
our concern here. We simply assume that the trace anomaly has two
components, the dominant ``hard" gluonium component and the
subdominant ``soft" quarkonium component. In a spirit close to
that taken by Beane and van Kolck~\cite{BK94}, we assume that in
matter, the hard component gradually decouples with increasing
density and in the so-called dilaton limit reached at the critical
point, only the quarkonium remains active. Thus the ``soft"
component is instrumental in describing the chiral properties in
dense systems~\footnote{In temperature driven chiral restoration,
roughly half of the gluon condensate $\la G^2\ra$ decondenses
across the phase transition~\cite{miller}. Since the gluonium
contribution dominates at zero temperature, this means that part
of the gluonium contribution must also melt at the transition, not
just the quarkonium part which is small at $T=0$. We have no clear
idea as to how to unravel this subtlety but we believe this not to
be crucial for the qualitative argument we are developing here.}.
In order to represent the degree of our ignorance, we consider
three possible scenarios with the mass of the scalar $m_\chi =$
550, 720 and 1000 MeV. The results obtained from the three do not
differ qualitatively, but the densities at which the interesting
phenomena take place do change substantially.

The physics we analyze in this new model is closely related with
the one presented in our earlier work~\cite{LPMRV03}. However the
details are modified dramatically from the previous work. As
before, the skyrmion matter we consider has two phases: a
low-density phase, which we ultimately describe by an
inhomogeneous phase, and a high-density phase, which is described
by a CC half-skyrmion crystal. In the former phase which we
simulate up to at most a few times nuclear matter density by an
inhomogeneous structure -- the closest we can get at present to a
Fermi liquid phase -- the pions in the medium are massive and
their dynamics determined by an effective theory where chiral
symmetry is spontaneously broken, while the scalar has a large
mass and couples strongly to the pions. The parameters governing
the phase transition, $\langle\sigma\rangle$ and $\chi_0$, are
non-vanishing. In the dense phase, on the contrary, skyrmion
matter becomes a stable crystal where the two parameters vanish
simultaneously at the same density. The pions become massless and
their dynamics is governed by a chirally restored theory, while
the scalar remains with a small but finite mass decoupled from the
pions. The fact that both parameters vanish at the same density
confirms the link between chiral symmetry restoration and the
``soft" component of the $\chi$ field. The scale anomaly is
maintained by the background field and the ``hard" component,
which decouples from the pions and describes scalar gluonium by
itself. Our model realizes specifically the ``lifting" of the
radius-field scenario of Chanfray {\it et al}~\cite{CEG01} and
provides a more precise meaning to the scaling behavior proposed
in \cite{BR91}. The field $\chi_0$ defines the radius of the
chiral circle for the in-medium pions and therefore when it is
non-vanishing their ground state is degenerate and chiral symmetry
is spontaneously broken. At the phase transition the chiral circle
abruptly shrinks to zero and chiral symmetry is restored. In our
model the background fields, $\chi_0$ and $(\sigma, \vec{\pi})$,
representing skyrmion matter play a dominant role in this
``lifting" process.

The model enables one to calculate leading corrections to
``Brown-Rho" scaling~\cite{BR91}. The in-medium $f_\pi$ decreases
with density at approximately the same rate as in \cite{BR91} for
low densities. However in our model, two mechanisms participate in
this decrease, the changing of $\chi_0$ and the deformation of the
skyrmion background fields. As mentioned, the main difference with
respect to our previous work of \cite{LPMRV03} is that here
$f_\pi$ vanishes in the dense phase whereas it does not in
\cite{LPMRV03}. This vanishing is solely associated with the
scalar field.

The masses of the pions and the scalar meson decrease with density
in a very characteristic way. The pion mass does not change much
over the range of density relevant to our consideration. Indeed
for low density its mass basically remains constant. Not so for
the scalar, whose mass decreases rapidly, so much so that close to
the phase transition it becomes nearly degenerate with the pion.
This phenomenon may be related to the ``mended symmetry" in the
dilaton limit of the Beane-van Kolck scenario~\cite{BK94} and
provides a support to our understanding of the behavior of the
$\chi$ field. However the scalar mass remains finite at the phase
transition and in the dense phase, while the pion mass vanishes
identically. The reason for the non-vanishing of the scalar mass
is associated to the scale anomaly contribution from the
background fields. The corresponding term is small. However when
all the remaining terms vanish according to the Beane-van Kolck
scenario, the small term becomes relevant. In the true dilaton
limit the scalar should also become massless and would become the
$\sigma$-meson of the sigma model Lagrangian. However the
background obstructs this limit through its contribution to the
scale anomaly and gives the $\sigma$ a non-vanishing mass.

We have also calculated the decay width of the scalar in the
medium, and we realize that at the phase transition the decay
width vanishes, signalling the consolidation of the scalar as a
hadronic stable particle and a possible manifestation of chiral
restoration as discussed in the literature~\cite{HK01,F03}.

We now make a brief remark on the implications of this work on the
parametric dependence of an effective field theory matched to QCD
discussed in the introduction. What we have computed is the
density dependence of the ratio $f_\chi^*/f_\chi$ and the response
to density of the background skyrmion. They should provide density
dependence to $all$ variables of interest in the model. Since we
have no vector degrees of freedom explicitly present in the model,
we cannot address directly how the gauge coupling $g$ scales.
However we can think of implementing the vectors in a manner
consistent with HLS in which case, their scaling is likely
governed by the scaling of the the $\chi_0$ field as first
proposed in \cite{BR91} with corrections coming from the change in
the background skyrmion. Thus it is highly likely that we will
recover at the chiral transition the limit $g\rightarrow 0$
together with $m_\rho^*/m_\rho\rightarrow 0$ as in BR scaling of
\cite{BR91}.

An important feature of the description of the medium properties,
discussed in the literature \cite{SS02a} - \cite{HKRS02}, is the
behavior of the so called pion velocity. This quantity denoted
$v_\pi$ is given by the ratio of the space part over the time part
of the pion decay constant $f_\pi^s/f_\pi^t$ and characterizes how
different effective schemes approach $QCD$ around the chiral phase
transition as proposed in \cite{HKRS02}. In hot/dense matter the
dynamics breaks Lorentz invariance and hence the different
components could differ in general. For instance, when chiral
restoration is induced by temperature, there can take place two
drastically different phenomena. If one assumes that in the
vicinity of chiral restoration the only relevant degrees of
freedom are pions, then the pion velocity is found~\cite{SS02a} to
vanish $v_\pi\rightarrow 0$ as $T\rightarrow T_c$. However if
there are other light degrees of freedom that can enter in the
chiral transition as, e.g., in the hidden local symmetry theory of
Harada and Yamawaki~\cite{HY:PR} where the light $\rho$-mesons
with the vector manifestation play a crucial role, the
prediction~\cite{HKRS02} is that the velocity approaches 1 as the
critical temperature is approached. There can be a small deviation
from 1 if one takes into account Lorenz-breaking terms in medium
at the scale at which the effective field theory (HLS/VM) and QCD
correlators are matched. In our present scheme where we are
concerned with dense matter, the situation is quite analogous to
the case of HLS/VM. At the leading order of pion fluctuation in
the background of the skyrmion matter, the pion velocity is 1.
This is because the influence of the background at this order is
principally from the scalar field which is Lorentz-scalar. The
deviation from 1 can come at the next order where the pion
interacts with the static soliton background. It turns out that
the deviation from 1 in $v_\pi$ remains small at least in
perturbation theory, resembling the case of HLS/VM at $T\approx
T_c$. The details will be given in a subsequent
publication~\cite{LPRV03}.

There are several important issues with the model that remain to
be addressed. First of all, we treated the normal matter to be an
inhomogeneous matter to which the FCC crystal state collapses.
This is not quite the Fermi-liquid state we know nuclear matter
should be in. To obtain a Fermi-liquid state from skyrmion matter,
we need, at the very least, to quantize the collective degrees of
freedom of the skyrmion matter. This is being worked out at
present. The solution to this problem will provide us a more
realistic information on the intrinsic dependence on the
background than what we have done here.

The phase that possesses Wigner symmetry at high density is a
half-skyrmion CC crystal and it is most likely a solid. It is not
clear whether this state -- when the color gauge fields are
suitably implemented -- is color-superconducting as predicted by
QCD at asymptotic density~\cite{wilczek} or a precursor to color
superconductivity. Ignoring this, suppose that such a solid state
is stabilized in compact stars. An interesting question is what
the solid state implies for the properties of dense stars and
their possible observable signatures~\cite{xu}. A related question
is whether this state is different from quark stars or other
exotic stars that have been discussed in the literature. This is
an issue that may be confronted with the phenomenology of compact
stellar systems.

\section*{Acknowledgments}
Hee-Jung Lee, Byung-Yoon Park and Vicente Vento are grateful for
the hospitality extended to them by KIAS, where this work was
initiated. This work was partially supported by grants
MCyT-BFM2001-0262 and GV01-26~(VV), KOSEF Grant
R01-1999-000-00017-0~(BYP), M02-2002-000-10008-0~(HJL)

\end{document}